\documentclass[twocolumn,aps,showpacs,preprintnumbers,amsmath,amssymb]{revtex4}
 \usepackage{graphicx}

\begin{document}
  \title{Competition between Electromagnetically Induced Transparency and Raman Processes}
  \author{G. S. Agarwal}
  \altaffiliation{On leave of absence from Physical Research
  Laboratory, Navrangpura, Ahmedabad - 380 009, India.}
  \author{T. N. Dey}
  \email{tarak.dey@okstate.edu}
  \affiliation{Department of Physics, Oklahoma State University,
  Stillwater, Oklahoma 74078, USA}%
  \author{Daniel J. Gauthier}
  \affiliation{Department of Physics, Duke University, Durham, North Carolina, 27708, USA}
  \date{\today}
  \begin{abstract}
 We present a theoretical formulation of  competition among
 electromagnetically induced transparency (EIT) and Raman
 processes. The latter become important when the medium can no
 longer be considered to be dilute. Unlike the standard formulation
 of EIT, we consider all fields applied and generated as
 interacting with both the transitions of the $\Lambda$ scheme. We
 solve Maxwell equations for the net generated field using
 a fast-Fourier-transform technique and obtain predictions for
 the probe, control and Raman fields. We show how the intensity of the
 probe field is depleted at higher atomic number densities
 due to the build up of multiple Raman fields.
 \end{abstract}


 \maketitle
 Multilevel atomic system interacting with several electromagnetic
 fields can give rise to variety of phenomena that depend
 on the strength and detunings of the fields.  Often, the various
 processes compete with each other, whereby some processes
 are suppressed or interference between processes renders
 the medium transparent to the applied fields \cite{Gauthier_JCP_1993}.
 Well known examples include competition between
 third-harmonic generation and
 multiphoton ionization \cite{Tewari_PRL_86}, and
 four-wave mixing and two-photon
 absorption \cite{Boyd_PRL_85,Agarwal_PRL_86}. Very recently,
 Harada \textit{et al.} demonstrated experimentally that
 stimulated Raman scattering can disrupt electromagnetically
 induced transparency (EIT), where the incident probe beam
 is depleted and new fields are generated via Raman
 processes \cite{Motomura_PRA_06}. The disruption of EIT
 is important to understand because it may degrade
 the performance of EIT-based applications, such as
 optical memories and  buffers, and magnetometers.
 In this paper we present a theoretical
 formulation that enables us to study the competing EIT and
 various orders of Raman processes to all orders in the
 applied and generated fields.

\begin{figure}
 \scalebox{0.52}{\includegraphics{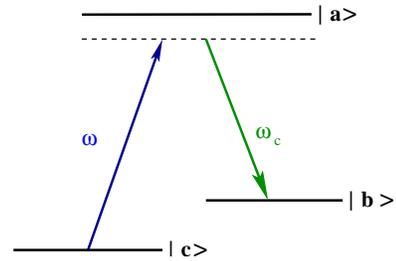}}\vspace{0.2in}
 \caption{\label{Fig1}(Color online) A schematic diagram of a three-level atomic
 system with energy
 spacing $\hbar\omega_{bc}$ between two ground states $|c\rangle$ and
 $|b\rangle$. The control field with frequency $\omega_c$ and
 probe field with frequency $\omega$ act on the atomic transitions
 $|a\rangle\leftrightarrow|b\rangle$ and $|a\rangle\leftrightarrow|c\rangle$, respectively.}
 \end{figure}

 The standard treatment of EIT \cite{Harris_PT_97} is based on the scheme of Fig.~\ref{Fig1},
 where the atoms in the state $|c\rangle$ interact with a probe
 field of frequency $\omega$.
 \begin{figure}
 \scalebox{0.52}[.52]{\includegraphics{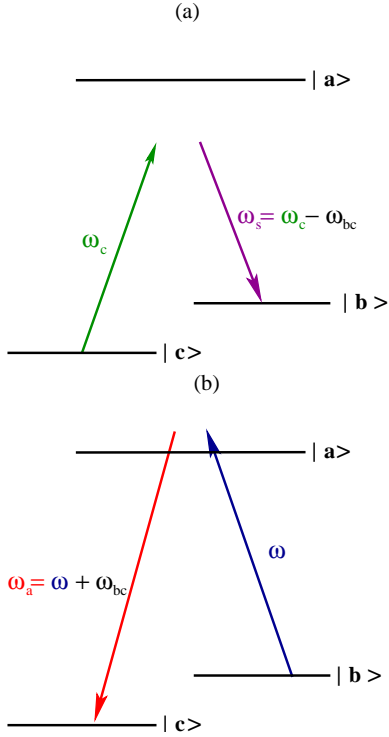}}\vspace{.5in}
 \caption{\label{Fig2}(Color online) Diagrammatic explanation of the (a)
 Stokes and (b) anti-Stokes processes. The intermediate state is denoted by $|a\rangle$.}
 \end{figure}
 \begin{figure}
 \scalebox{0.52}{\includegraphics{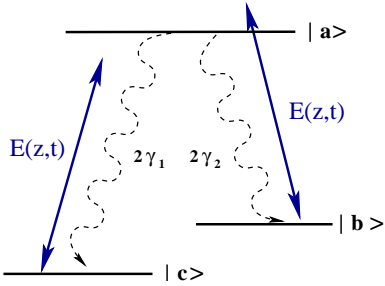}}
 \caption{\label{Fig3}(Color online) Three-level $\Lambda$
 system interacting with the space-time dependent field
 $E(z,t)$ on both the optical transitions.}
 \end{figure}
 A control field of frequency
 $\omega_c$ interacts on the unoccupied transition
 $|a\rangle\leftrightarrow|b\rangle$. The probe and the control
 fields are tuned such that
 \begin{equation}\label{EIT}
 \omega - \omega_c = \omega_{bc}.
 \end{equation}
 This results in no absorption of the probe field provided the
 coherence $\rho_{bc}$ has no decay. This treatment assumes that the
 frequency separation $\omega_{bc}$ is so large that the
 interaction of the control field $\omega_c~(\textrm {probe field}~\omega)~$
 with the transition
 $|a\rangle\leftrightarrow|b\rangle~(|a\rangle\leftrightarrow|c\rangle)~$can
 be ignored.  At high atomic number densities or for strong fields, this
 approximation no longer holds, which is the situation we consider here.

 At higher densities, Raman processes start becoming
 important \cite{Raman,RamanK}, such as those shown in the Fig~\ref{Fig2},
 for example.  The Raman generation of the fields at $\omega_c-\omega_{bc}$,
 $\omega+\omega_{bc}$ can further lead to newer frequencies like
 $\omega_c-2\omega_{bc}$. In order to account for the Raman processes,
 we write the electromagnetic field acting on both the transitions as
 \begin{equation}\label{Field}
 E(t)={\cal E}(t)e^{-i\omega_c t},
 \end{equation}
 where ${\cal E}(t)$ denotes the net generated field. At the input
 face of the medium ${\cal E}(t)$ has two components to account
 for both control and probe fields
 \begin{equation}\label{netField}
 {\cal E}(t)={\cal E}_c + {\cal E}_p~e^{-i(\omega-\omega_c)t}.
 \end{equation}
 Under the Raman-resonance condition (\ref{EIT}),
 we expect ${\cal E}(t)$ to have the structure
 \begin{equation}\label{geField}
 {\cal E}(t)=\sum{\cal E}^{(n)} e^{-in\omega_{bc} t}.
 \end{equation}
 Thus, ${\cal E}^{(-1)}$ gives the strength of the Stokes
 process of the Fig.~\ref{Fig2}(a); ${\cal E}^{(+2)}$ gives the strength of
 the process of the Fig.~\ref{Fig2}(b); and ${\cal E}^{(+1)}$ describes the
 changes in the probe field. For low atomic number densities, we
 expect the usual
 results and therefore ${\cal E}^{(+1)}\approx{\cal E}$ and ${\cal E}^{(0)}\approx{\cal
 E}_c$.

 To calculate the net generated field ${\cal E}(t)$ for arbitrary
 atomic number density, we
 have to solve the coupled Maxwell and density matrix
 equations. We consider now the situation as shown schematically
 in the Fig.~\ref{Fig3}.
 The applied field $E(z,t)$ couples the
 excited state $|a\rangle$ to both ground states $|b\rangle$ and $|c\rangle$ .
 Here $2\gamma$'s represents rates of spontaneous
 emission. In a frame rotating with the frequency $\omega_c$ the
 density matrix equation for the atomic system are given by
 \begin{eqnarray}\label{density}
 \dot{\rho}_{aa}&=&i \Omega_{\cal E} \left(\rho_{ba}+
 \rho_{ca}\right)
 -i \Omega_{\cal E}^{*} \left(\rho_{ab}+\rho_{ac}\right)-4\gamma\rho_{aa}~,\nonumber\\
 \dot{\rho}_{bb}&=& i \Omega_{\cal E}^{*} \rho_{ab} - i \Omega_{\cal E} \rho_{ba}
 + 2\gamma\rho_{aa}~,\nonumber\\
 \dot{\rho}_{ab}&=&-[2\gamma-i\Delta_c]\rho_{ab}+i \Omega_{\cal E}(\rho_{bb}-\rho_{aa})
 +i \Omega_{\cal E} \rho_{cb}~,\\
 \dot{\rho}_{ac}&=&-[2\gamma-i(\Delta_c-\omega_{bc})]\rho_{ac}
 +i \Omega_{\cal E} \rho_{bc}+i \Omega_{\cal E}(\rho_{cc}-\rho_{aa})~,\nonumber\\
 \dot{\rho}_{bc}&=&-(\Gamma_{bc}+i\omega_{bc})\rho_{bc}+i \Omega_{\cal E}^{*}\rho_{ac}
 -i \Omega_{\cal E} \rho_{ba}~,\nonumber
 \end{eqnarray}
 \begin{figure}
 \includegraphics[width=0.46\textwidth]{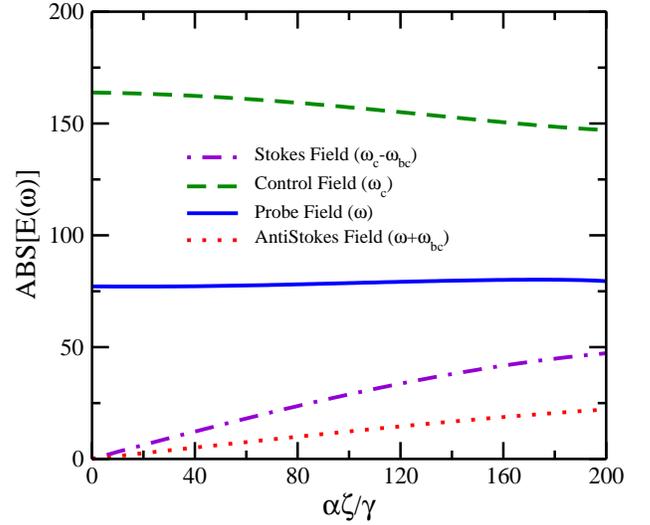}
 \caption{\label{Fig4}(Color online) Amplitudes of different Fourier
 components of the net generated field as a function of the atomic density
 of the medium.
 The normalized propagation length $\alpha\zeta/\gamma=200$
 is equivalent to  actual length of the medium L=7.13 cm with an atomic density of $n=10^{10}$ atoms/cm$^3$.
 The other parameters of the above graph are chosen as:
 input Rabi frequency $\Omega_{\cal E} =0.5\gamma$,
 ${\cal E}_p/{\cal E}_c=0.5$, $\Delta_c=0.0$,
 $\gamma=9.475\times 10^6$, $\Gamma_{bc}=0.0$, $\omega_{bc}=100\gamma$ and $\lambda=766.4$ nm.}
 \end{figure}
 where the detuning $\Delta_c$ and the space and time dependent Rabi
 frequency $\Omega_{\cal E}$ of the generated fields are defined by
 \begin{equation}\label{Rabi}
 \Delta_c= \omega_c-\omega_{ac};~~~\Omega_{\cal E}(z,t)=\frac{\vec{d}\cdot\vec{{\cal
 E}}}{\hbar}.
 \end{equation}
 For simplicity, we have assumed $\vec{d}_{ab}=\vec{d}_{ac}=\vec{d}$. The elements $\rho_{ac}$ and $\rho_{ab}$ in the original frame
 can be
 obtained by multiplying the solution of Eqs.~(\ref{density})
 by $e^{-i\omega_c t}$.
 The induced polarization $\vec{\cal P}$ is given by
 \begin{equation}\label{Polarization}
 \vec{\cal P}=\left(\vec{d}\rho_{ab}+\vec{d}\rho_{ac}\right)e^{-i\omega_c t}.
 \end{equation}
 The Maxwell equations in the slowly varying envelope approximation
 lead to the following equation for the generated field
 \begin{equation}\label{Maxwell}
 \left(\frac{\partial \Omega_{\cal E} }{\partial z}+ \frac{\partial \Omega_{\cal E}}{\partial
 ct}\right)=i\frac{\alpha}{2}
 \left(\rho_{_{ac}}+\rho_{_{ab}}\right),
 \end{equation}
 where $\alpha$ is given by
 \begin{equation}\label{coupling}
 \alpha=3\lambda^2n\gamma/4\pi ,
 \end{equation}
 and $n$ is the atomic density. The coupled
 equations (\ref{density}) and (\ref{Maxwell}) are solved in the
 moving coordinate system
 \begin{equation}\label{window}
 \tau=t-\frac{z}{c};~\zeta=z.
 \end{equation}

 We have numerically solved the coupled set of equations when all the
 atoms are
 initially in the state $|c\rangle$ and when the fields at the
 input face of the medium are given by (\ref{netField}). We
 calculate ${\cal E}(l,\tau)$ and do a fast Fourier transform to
 obtain the different Fourier components of the field at the output
 face of the medium.
 This procedure enables us to find how the probe and control fields
 evolve and determine when the Raman processes become important.
 In the simulations, we have used parameters
 that are appropriate for $^{39}$K vapor \cite{RamanK} to avoid numerical
 problems.  In this situation, the spontaneous decay rate of the
 excited state $|a\rangle$
 $4\gamma=3.79\times 10^{7}$ rad/s and the wavelength for the
 ground state $|c\rangle$
 to excited state $|a\rangle$ transition $\lambda=766.4$ nm.
 \begin{figure}
 \scalebox{0.48}{\includegraphics{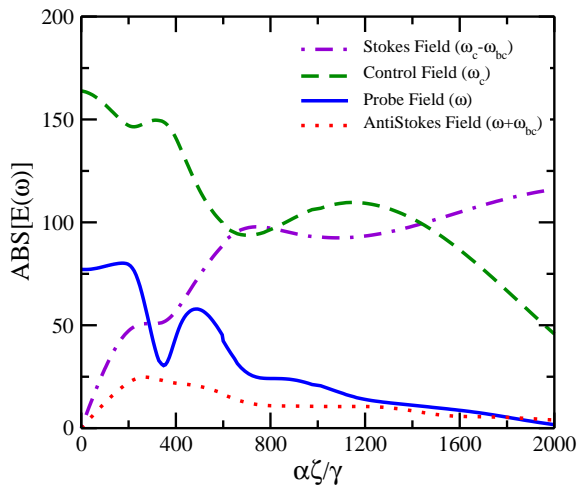}}\vspace{0.4in}
 \caption{\label{Fig5}(Color online) The spectral amplitudes of different fields are plotted against the atomic density of the medium.
 The normalized propagation length $\alpha\zeta/\gamma=2000$
 is equivalent to actual length of the medium L=7.1 cm when the atomic density $n=10^{11}$ atom/cm$^3$.
 The other parameters are chosen as: input
 Rabi frequency $\Omega_{\cal E} =0.5\gamma$,
 ${\cal E}_p/{\cal E}_c=0.5$, $\Delta_c=0.0$, $\Gamma_{bc}=0.0$ and $\omega_{bc}=100\gamma$.}
 \end{figure}

 \noindent{\bf EIT Vs Raman Processes} In this section, we present the
 results of numerical calculations. In Fig.~\ref{Fig4}, we show
 result for the low-density regime.
 In this region, we notice almost no change in the probe
 field and thus EIT dominates. It is also seen that the
 Raman processes slowly start to build up, leading to
 the drop in the control field amplitude.

 We next consider the high-density regime, as shown in Fig.~\ref{Fig5}.
 This is the region when multiple Raman processes build up
 significantly \cite{footnote1}.
 Our numerical results are in broad agreement with the observations of Harada
 {\it et al.}\cite{Motomura_PRA_06} where they observe
 the depletion of the probe field
 and the generation of the Stokes field at $(\omega_c-\omega_{bc})$.
 In particular, we see in fig.~\ref{Fig5} that the generation of
 radiation at $(\omega_c-\omega_{bc})$ is very
 important and the probe beam is depleted.  We also notice a new
 feature - the probe exhibits some oscillatory character before
 dying out. This oscillation is due to the fact that
 any population that is transferred to
 the state $|b\rangle$ can produce a field at the probe frequency
 via the Raman process. When this happens, the control field amplitude
 falls.

\begin{figure}
 \scalebox{0.48}{\includegraphics{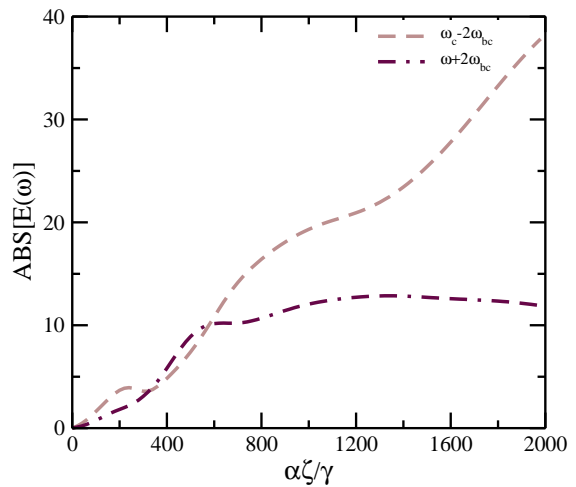}}\vspace{0.2in}
 \caption{\label{Fig6}(Color online) The amplitudes of hyper Raman
 components at frequencies $(\omega_c-2\omega_{bc})$, $(\omega+2\omega_{bc})$
 as a function of the atomic density.
 All parameters are same as in Fig.~(\ref{Fig5}).}
 \end{figure}

\begin{figure}\vspace{0.3in}
 \scalebox{0.44}{\includegraphics{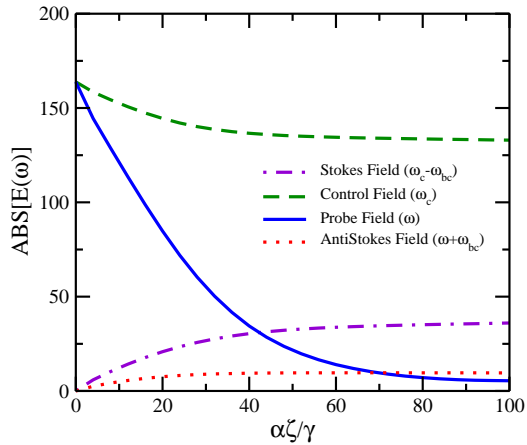}}\vspace{0.05in}
 \caption{\label{Fig7}(Color online) The amplitudes of Stokes, control, probe and anti-Stokes
 fields as a function of the atomic number density in the
 presence of a buffer
 gas. Here, we have scaled the amplitudes of
 probe, Stokes and anti-Stokes fields by a factor of two.
 The other parameters are chosen as: density $n=10^{10}$ atom/cm$^3$,
 input Rabi frequency $\Omega_{\cal E} =0.5\gamma$,
 ${\cal E}_p/{\cal E}_c=0.5$, $\Delta_c=0.0$, $\Gamma_{bc}=0.01\gamma$ and $\omega_{bc}=100\gamma$.
 A nonzero value of $\Gamma_{bc}$ accounts for the buffer gas.}
 \end{figure}

 In Fig.~\ref{Fig6}, we show the build up of several hyper-Raman processes.
 The effect of a buffer gas on the generated field is
 shown in Fig.~\ref{Fig7}, where it is seen that the
 amplitude of the probe field depletes faster in the presence of buffer
 a gas.
 On comparison of Fig.~\ref{Fig5} and
 Fig.~\ref{Fig7}, we see that the amplitudes of the probe field
 and the generated Raman field become equal at $\alpha\zeta/\gamma=272$
 (without buffer gas) and 84 (with buffer gas). This is in agreement
 with the observation in Ref.\cite{Motomura_PRA_06}.

 In conclusion, we have investigated competition
 between electromagnetically induced transparency and Raman processes in a $\Lambda$ system
 due to the cross talk among the optical transitions. We have
 demonstrated that the EIT-induced probe spectrum is very
 pronounced in comparison to the higher order Raman sidebands
 for a low atomic number density. However, the
 generated Raman fields become dominant for an atomic
 number density that is only ten times higher.

 \end{document}